\documentclass{iopconfser}
\usepackage{orcidlink}
\usepackage{amsmath,amssymb}
\usepackage[numbers,sort&compress]{natbib}

\usepackage{etoolbox}
\newtoggle{checklength}
\toggletrue{checklength} 

\usepackage{acronym}

\begin{document}

\title{Hunting for new glitches in LIGO data using community science}

\author{E~Mackenzie$^{1}$\orcidlink{0009-0001-5967-7599},
C~P~L~Berry$^{1}$\orcidlink{0000-0003-3870-7215}, 
G~Niklasch$^{2,3}$, 
B~Téglás$^{2}$, 
C~Unsworth$^{2}$, 
K~Crowston$^4$\orcidlink{0000-0003-1996-3600}, 
D~Davis$^{5}$\orcidlink{0000-0001-5620-6751} and 
A~K~Katsaggelos$^{6,7}$\orcidlink{0000-0003-4554-0070}}

\iftoggle{checklength}{
\affil{$^1$Institute for Gravitational Research, University of Glasgow, Kelvin Building, University Ave., Glasgow, G12 8QQ, United Kingdom} 

\affil{$^2$Gravity Spy} 

\affil{$^3$ConSol Consulting \& Solutions Software GmbH, St.-Cajetan-Straße 43, D-81669, Munich, Germany} 

\affil{$^4$School of Information Studies, Syracuse University, Hinds Hall, Syracuse, 13210, NY, USA}

\affil{$^5$LIGO Laboratory, California Institute of Technology, Pasadena, CA 91125, USA}

\affil{$^6$Department of Electrical and Computer Engineering, Northwestern University, Evanston, IL 60208, USA}

\affil{$^7$NSF-Simons AI Institute for the Sky (SkAI), Chicago, IL 60611, USA}
}

\email{christopher.berry.2@glasgow.ac.uk}

\begin{abstract}
Data from ground-based gravitational-wave detectors like LIGO contain many types of noise. 
Glitches are short bursts of non-Gaussian noise that may hinder our ability to identify or analyse gravitational-wave signals. 
They may have instrumental or environmental origins, and new types of glitches may appear following detector changes. 
The Gravity Spy project studies glitches and their origins, combining insights from volunteers on the community-science Zooniverse platform with machine learning. 
Here, we study volunteer proposals for new glitch classes, discussing links between these glitches and the state of the detectors, and examining how new glitch classes pose a challenge for machine-learning classification. 
Our results demonstrate how Zooniverse empowers non-experts to make discoveries, and the importance of monitoring changes in data quality in the LIGO detectors. 
\end{abstract}

\section{Detector characterisation and glitches}

Analysing gravitational-wave data from detectors such as the Laser Interferometer Gravitational-Wave Observatory (LIGO)~\cite{TheLIGOScientific:2014jea} requires understanding the properties of noise~\cite{LIGOScientific:2019hgc}. 
Analysis is complicated by the presence of \emph{glitches}, short non-Gaussian noise transients. 
The most infamous glitch may be the one that overlapped GW170817 in LIGO Livingston data, which needed to be subtracted from the data before source properties could be inferred~\cite{LIGOScientific:2017vwq,Pankow:2018qpo}. 
Glitches come in a wide variety of forms reflecting the diversity of their origins~\cite{TheLIGOScientific:2016zmo,Nuttall:2018xhi,Glanzer:2022avx}, and different glitch types may have different impacts on analyses~\cite[e.g.,][]{Davis:2020nyf,Pankow:2018qpo,Powell:2018csz,Macas:2022afm,Kwok:2021zny}. 

There are a variety of detector-characterisation tools used to study glitches~\cite{Davis:2021ecd,LIGO:2024kkz}. 
These analyse the main gravitational-wave strain channel, as well as auxiliary channels that monitor the internal state of the instrument as well as its surroundings~\cite{Nguyen:2021ybi}; e.g., Omicron identifies trigger times with excess power in the data stream~\cite{Robinet:2020lbf}, and these triggers can be used in follow-on analysis such as Hveto~\cite{Smith:2011an}, which looks for correlations in triggers from gravitational-wave and auxiliary channels. 
Ideally, we would either identify the origin of glitches such that they can be eliminated~\cite[e.g.,][]{Soni:2020rbu,Berger:2023pyq}, or characterise their properties such that their impact on data analysis can be ameliorated~\cite[e.g.,][]{Davis:2022ird,Merritt:2021xwh,Udall:2022vkv,Tolley:2023umc,Malz:2025xdg}.

The Gravity Spy project studies glitches using a combination of community (or citizen) science and machine learning~\cite{Zevin:2016qwy,Zevin:2023rmt}. 
The machine-learning algorithm efficiently sorts large numbers of glitches into different classes~\cite{Bahaadini:2018git,Wu:2024tpr}. 
However, it can struggle when new types of glitches appear. 
The community-science aspect of the project is run through the Zooniverse platform, where volunteers classify spectrograms~\cite{Chatterji:2004qg} of gravitational-wave and auxiliary-channel data.%
\footnote{Gravity Spy Zooniverse project \href{http://gravityspy.org/}{gravityspy.org}.}
The volunteers can make use of Zooniverse infrastructure, such as the Talk forum to collaborate in their efforts. 
The Gravity Spy project provides a system of levels, to help train volunteers in classification~\cite{Jackson:2020}, and tools such as a Similarity Search~\cite{Coughlin:2019ref} to empower them to investigate the data. 
While most of the $37,000$ volunteers classify glitches into known classes, a few advanced volunteers search for new classes.
The work of the volunteers can help us identify new features in the data, and build training sets to update the machine-learning algorithm~\cite{Soni:2021cjy,Zevin:2023rmt}.

Since the launch of Gravity Spy in 2016, collaboration between Zooniverse volunteers and LIGO detector-characterisation experts has resulted in several new glitch classes being added to the project~\cite{Soni:2021cjy,Glanzer:2022avx}. 
Here, we review two volunteer proposals for new glitch classes. 

\section{Volunteer glitch proposals}

\begin{figure}
    \centering
    \includegraphics[width=0.78\columnwidth]{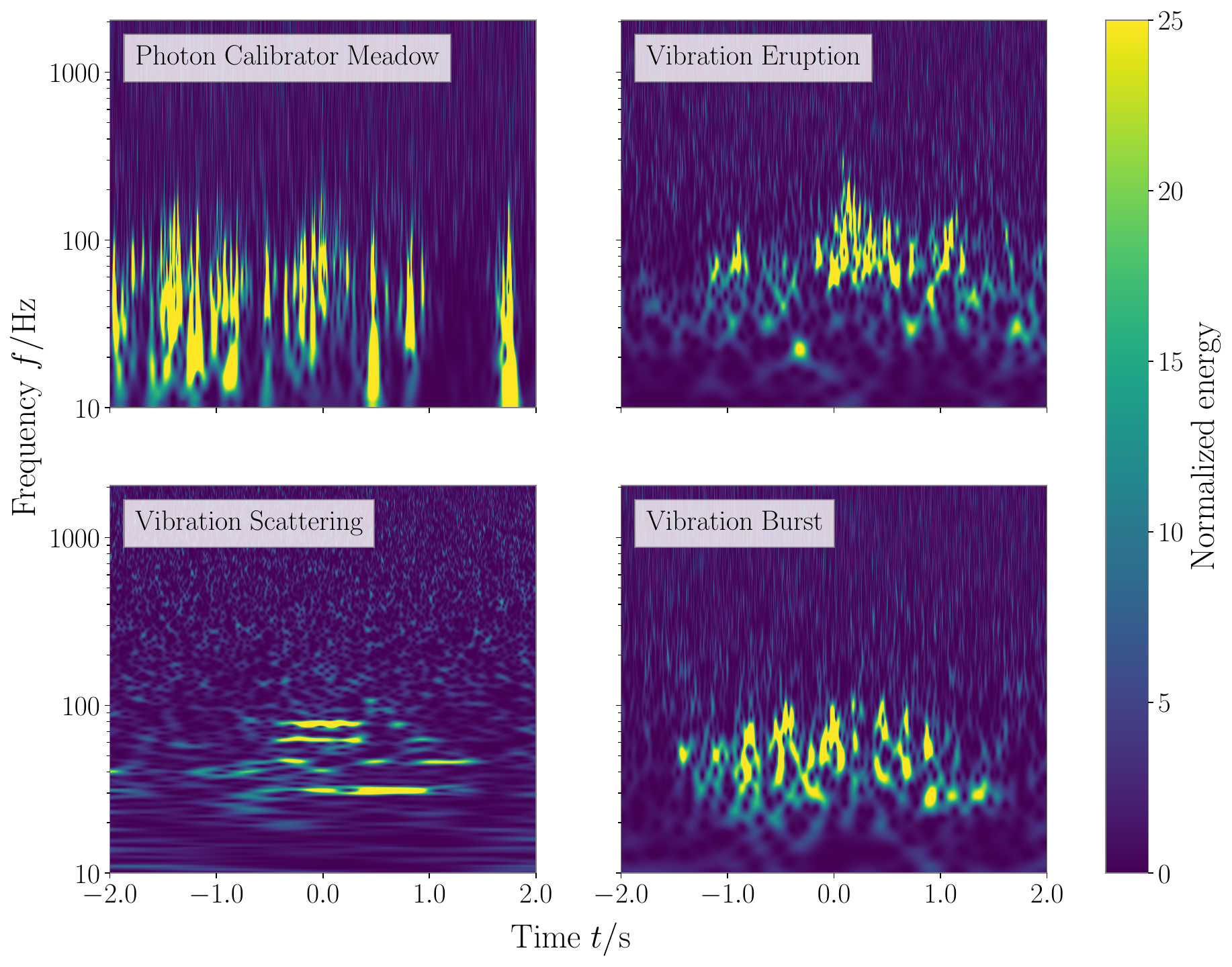}
    \caption{Example time--frequency spectrograms~\cite{Chatterji:2004qg} for the proposed Photon Calibrator Meadow class and the three potential subclasses of the proposed Vibration class. 
    Similar spectrograms, showing time windows of $0.5~\mathrm{s}$,  $1~\mathrm{s}$, $2~\mathrm{s}$ and $4~\mathrm{s}$ are examined by volunteers on Zooniverse~\cite{Zevin:2016qwy,Zevin:2023rmt}. 
    These examples are all taken from LIGO Livingston data.}
    \label{fig:spectrogram}
\end{figure}

Proposals for new classes are made using the Zooniverse Talk forum, and follow a template~\cite{Soni:2021cjy}.%
\footnote{Process for making new glitch classes official \href{https://www.zooniverse.org/projects/zooniverse/gravity-spy/talk/762/497614}{www.zooniverse.org/projects/zooniverse/gravity-spy/talk/762/497614}.}
The template asks for information such as a proposed name, a prototypical example, a description, a list of hashtags associated with subjects (used by volunteers to collate interesting glitches), and links to collections of examples of the proposed class. 
Proposals often include information from investigations done by experienced volunteers. 
We will discuss two proposals: Photon Calibrator Meadow (Section~\ref{sec:pcal}) and Vibration (Section~\ref{sec:vibration}); example spectrograms from these proposals are shown in Figure~\ref{fig:spectrogram}.

\subsection{Photon Calibrator Meadow}\label{sec:pcal}

This glitch appears as a field of many flame-like transients below $\sim256~\mathrm{Hz}$. 
The proposal was made 2 November 2024, with the examples from LIGO Livingston data on 18 August 2024.\footnote{L1 O4b Photon calibrator meadow \href{https://www.zooniverse.org/projects/zooniverse/gravity-spy/talk/762/3491179}{www.zooniverse.org/projects/zooniverse/gravity-spy/talk/762/3491179}.}  
Drawing upon public logs~\cite[e.g.,][]{alog:72717,alog:72799,alog:72854,alog:72867}, the volunteers hypothesised that the cause was related to an issue with the photon-calibrator system~\cite{Karki:2016pht} reported on that day.

Examples from this proposed class are primarily classified as Fast Scattering (Crown) or Extremely Loud by the machine-learning algorithm. 
Extremely Loud is an umbrella class for loud glitches without a single origin~\cite{Glanzer:2022avx}, so sorting into this class is not problematic, but significant contamination of the Fast Scattering class, which is caused by scattered light linked to ground motion~\cite{Soni:2021cjy,Soni:2023kqq,Glanzer:2023hzf}, would create difficulties for studies using the classifications. 

Our investigation shows that the glitches were mostly confined to a $<1~\mathrm{hr}$ period. 
This timing corresponded to a malfunction in the photon calibrator in the $Y$-arm of the detector.
Once the photon-calibrator issue was resolved, there were no further examples of this glitch. 

This proposed glitch class has a well-identified cause, potentially making it useful for detector-characterisation studies. 
However, it only occurred over the period with the now-fixed photon-calibrator issue, and has a low probability of reoccurring. 
Thus, we do not expect there to be further examples to classify (including this class would confuse future volunteer and machine-learning classifications), or a need for detector-characterization experts to study these glitches. 
Since it is a rare glitch class and easily identified by its time, there is not much concern about misclassifications contaminating the data set. 
Therefore, the Gravity Spy project will not add this glitch class yet, but we shall watch for reoccurrence.

Given that incidence of this class was well localized in time, a cause could be clearly identified in this case. 
The rarity of the glitch across the Gravity Spy data set (Zooniverse does not show glitches in strict chronological order)  demonstrates how well volunteers can identify novel features in large data sets.

\subsection{Vibration}\label{sec:vibration}

This glitch occupies mid-frequencies of $\sim30$--$200~\mathrm{Hz}$ and lasts for $\sim1$--$4~\mathrm{s}$. 
It does not have a simple shape, but often appears as a complicated and irregular network of peaks and horizontal lines. 
The proposal was made 24 May 2019, drawing upon examples from LIGO Livingston in the third observing run.\footnote{Vibration \href{https://www.zooniverse.org/projects/zooniverse/gravity-spy/talk/762/1009409}{www.zooniverse.org/projects/zooniverse/gravity-spy/talk/762/1009409}.} 
Similar examples are seen in the fourth observing run and in LIGO Hanford data.
The proposal includes potential subcategories of Vibration Eruption, which includes higher frequency ($\lesssim512~\mathrm{Hz}$) bursts, often at the start of the glitch; Vibration Burst, which lacks the higher frequency peaks, and Vibration Scattering, which shows horizontal structure similar to scattered-light glitches~\cite{Accadia:2010zzb,Ottaway:2012oce,Valdes:2017xce,Udall:2022vkv}. 

Glitch examples are primarily identified as Scattered Light (Slow Scattering)~\cite{Soni:2020rbu} or Scratchy~\cite{Glanzer:2022avx} by the machine-learning algorithm. 
This confusion reflects the similarity in morphologies. 
The different subcategories are classified differently: Vibration Eruption primarily as Scratchy, Scattered Light and Extremely Loud; Vibration Scattering as Scattered Light, and Vibration Bursts as Fast Scattering, Scratchy and Scattered Light.

Our investigations show that the three putative subclasses often occur in close proximity to each other (volunteers were only provided with the date), such that it makes sense to consider them all together. 
For example, Vibration Eruption was often observed a few seconds before Vibration Scattering. 
The time difference is similar to the $\sim12~\mathrm{s}$ sound-travel time across one of the arms.

Thunderstorms are a known source of noise for the detectors~\cite{Davis:2021ecd}. 
Depending upon the distance, direction and nature (a sharp crack verses a long rumble) of thunderclaps, they can be picked up by the detectors in different forms, with acoustic and ground-motion perturbations passing by different parts of the detector at different times. 
Storms are more common in Louisiana than Washington, so we expect related glitches to be more common in Livingston, as seen here. 
Thunderstorm glitches have been studied by detector-characterisation experts~\cite[e.g.,][]{Davis:2021ecd,Macas:2022afm}, and match those in the proposed class. 
Using auxiliary-channel information from microphones, a data-quality flag has been created to monitor thunderstorms~\cite{Davis:2022cmw}. 
Cross-referencing this flag, we find that $90\%$ of the compiled example Livingston glitches coincide with thunderstorms. 
The remaining glitches can potentially be attributed to anthropogenic noise sources, e.g., instances have been linked to times when liquid nitrogen was delivered~\cite{alog:47223}.

The volunteers have succeeded in identifying a glitch class with an environmental cause. 
As this glitch class is expected to be present in data from multiple observing runs and is of interest for detector-characterisation studies, it would be a useful addition to Gravity Spy, and the project will include it.

\section{Summary}

New glitches may arise or existing glitches may become more common following changes to the detector or their environments. 
It is therefore necessary to monitor the data quality of the LIGO detectors. 
We have presented a study of Gravity Spy volunteers' proposals for new glitches. 
We conclude that the Vibration class (as a single class) would be a useful addition to the Gravity Spy classes. 
The volunteers' collections can be drawn upon for retraining of the machine-learning classifier.

The volunteers have succeeded in identifying new features in the data, and used LIGO resources to investigate their origins. 
They were able to identify glitches from rare classes that only occurred for short periods of time.
This work illustrates how, given suitable tools and a training framework, members of the public can contribute to cutting-edge science. 

\iftoggle{checklength}{
\section*{Acknowledgments}
We thank the Gravity Spy volunteers, in particular Ruth Southern and Caspian D'silva, for all their efforts; we thank the rest of the Gravity Spy team for helpful feedback; we thank Jane Glanzer for thoughtful feedback on the manuscript, and we thank the LIGO Detector Characterization group for useful conversations.
This material is based upon work supported by National Science Foundation's (NSF's) LIGO Laboratory which is a major facility fully funded by the NSF, as well as Science and Technology Facilities Council (STFC) of the United Kingdom, the Max-Planck-Society (MPS), and the State of Niedersachsen/Germany for support of the construction of Advanced LIGO and construction and operation of the GEO\,600 detector. 
Additional support for Advanced LIGO was provided by the Australian Research Council. 
The authors are grateful for computational resources provided by the LIGO Laboratory and supported by 	
National Science Foundation Grants PHY-0757058 and PHY-0823459.
Gravity Spy is supported by NSF grants IIS-2106882, 2106896, 2107334, and 2106865. 
CPLB acknowledges support from STFC grant ST/V005634/1.
This document has been assigned LIGO document number LIGO-P2500468.
}

\bibliographystyle{iopart-num}
\bibliography{gravityspy}

@MISC{alog:47223,
   author =       {Taylor, G. and others},
   title =        {{aLIGO LLO Logbook}},
   howpublished =          "\href{https://alog.ligo-la.caltech.edu/aLOG/index.php?callRep=47223}{47223}",
   year = "2024"
}

@MISC{alog:72717,
   author =       {Heintze, M. and others},
   title =        {{aLIGO LLO Logbook}},
   howpublished =          "\href{https://alog.ligo-la.caltech.edu/aLOG/index.php?callRep=72717}{72717}",
   year = "2024"
}

@MISC{alog:72799,
   author =       {Bossilkov, V. and others},
   title =        {{aLIGO LLO Logbook}},
   howpublished =          "\href{https://alog.ligo-la.caltech.edu/aLOG/index.php?callRep=72799}{72799}",
   year = "2024"
}

@MISC{alog:72854,
   author =       {Kandhasamy, S.},
   title =        {{aLIGO LLO Logbook}},
   howpublished =          "\href{https://alog.ligo-la.caltech.edu/aLOG/index.php?callRep=72854}{72854}",
   year = "2024"
}

@MISC{alog:72867,
   author =       {Betzwieser, J. and others},
   title =        {{aLIGO LLO Logbook}},
   howpublished =          "\href{https://alog.ligo-la.caltech.edu/aLOG/index.php?callRep=72867}{72867}",
   year = "2024"
}

@article{TheLIGOScientific:2014jea,
    author = "Aasi, J. and others",
    collaboration = "LIGO Scientific Collaboration",
    title = "{Advanced LIGO}",
    eprint = "1411.4547",
    archivePrefix = "arXiv",
    primaryClass = "gr-qc",
    doi = "10.1088/0264-9381/32/7/074001",
    journal = "Class. Quant. Grav.",
    volume = "32",
    pages = "074001",
    year = "2015"
}

@article{TheLIGOScientific:2016zmo,
    author = "Abbott, B.P. and others",
    collaboration = "LIGO Scientific and Virgo Collaboration",
    title = "{Characterization of transient noise in Advanced LIGO relevant to gravitational wave signal GW150914}",
    eprint = "1602.03844",
    archivePrefix = "arXiv",
    primaryClass = "gr-qc",
    doi = "10.1088/0264-9381/33/13/134001",
    journal = "Class. Quant. Grav.",
    volume = "33",
    number = "13",
    pages = "134001",
    year = "2016"
}

@article{LIGOScientific:2017vwq,
    author = "Abbott, B. P. and others",
    collaboration = "LIGO Scientific and Virgo Collaboration",
    title = "{GW170817: Observation of Gravitational Waves from a Binary Neutron Star Inspiral}",
    eprint = "1710.05832",
    archivePrefix = "arXiv",
    primaryClass = "gr-qc",
    reportNumber = "LIGO-P170817",
    doi = "10.1103/PhysRevLett.119.161101",
    journal = "Phys. Rev. Lett.",
    volume = "119",
    number = "16",
    pages = "161101",
    year = "2017"
}

@article{LIGOScientific:2019hgc,
    author = "Abbott, Benjamin P and others",
    collaboration = "LIGO Scientific and Virgo Collaboration",
    title = "{A guide to LIGO\textendash{}Virgo detector noise and extraction of transient gravitational-wave signals}",
    eprint = "1908.11170",
    archivePrefix = "arXiv",
    primaryClass = "gr-qc",
    doi = "10.1088/1361-6382/ab685e",
    journal = "Class. Quant. Grav.",
    volume = "37",
    number = "5",
    pages = "055002",
    year = "2020"
}

@article{Accadia:2010zzb,
    author = "Accadia, T. and others",
    editor = "Ricci, Fulvio",
    title = "{Noise from scattered light in Virgo's second science run data}",
    doi = "10.1088/0264-9381/27/19/194011",
    journal = "Class. Quant. Grav.",
    volume = "27",
    pages = "194011",
    year = "2010"
}

@article{Bahaadini:2018git,
    author = "Bahaadini, S. and Noroozi, V. and Rohani, N. and Coughlin, S. and Zevin, M. and Smith, J. R. and Kalogera, V. and Katsaggelos, A.",
    title = "{Machine learning for Gravity Spy: Glitch classification and dataset}",
    doi = "10.1016/j.ins.2018.02.068",
    journal = "Info. Sci.",
    volume = "444",
    pages = "172--186",
    year = "2018"
}

@article{Berger:2023pyq,
    author = "Berger, B. K. and others",
    collaboration = "LIGO Instrument Science Collaboration",
    title = "{Searching for the causes of anomalous Advanced LIGO noise}",
    doi = "10.1063/5.0140766",
    journal = "Appl. Phys. Lett.",
    volume = "122",
    number = "18",
    pages = "184101",
    year = "2023"
}

@article{Chatterji:2004qg,
    author = "Chatterji, S. and Blackburn, L. and Martin, G. and Katsavounidis, E.",
    title = "{Multiresolution techniques for the detection of gravitational-wave bursts}",
    eprint = "gr-qc/0412119",
    archivePrefix = "arXiv",
    doi = "10.1088/0264-9381/21/20/024",
    journal = "Class. Quant. Grav.",
    volume = "21",
    pages = "S1809--S1818",
    year = "2004"
}

@article{Coughlin:2019ref,
    author = "Coughlin, S.B. and others",
    title = "{Classifying the unknown: discovering novel gravitational-wave detector glitches using similarity learning}",
    eprint = "1903.04058",
    archivePrefix = "arXiv",
    primaryClass = "astro-ph.IM",
    doi = "10.1103/PhysRevD.99.082002",
    journal = "Phys. Rev. D",
    volume = "99",
    number = "8",
    pages = "082002",
    year = "2019"
}

@article{Davis:2020nyf,
    author = "Davis, Derek and White, Laurel V. and Saulson, Peter R.",
    title = "{Utilizing aLIGO Glitch Classifications to Validate Gravitational-Wave Candidates}",
    eprint = "2002.09429",
    archivePrefix = "arXiv",
    primaryClass = "gr-qc",
    doi = "10.1088/1361-6382/ab91e6",
    journal = "Class. Quant. Grav.",
    volume = "37",
    number = "14",
    pages = "145001",
    year = "2020"
}

@article{Davis:2021ecd,
    author = "Davis, D. and others",
    collaboration = "LIGO Instrument Science Collaboration",
    title = "{LIGO Detector Characterization in the Second and Third Observing Runs}",
    eprint = "2101.11673",
    archivePrefix = "arXiv",
    primaryClass = "astro-ph.IM",
    reportNumber = "P2000495",
    doi = "10.1088/1361-6382/abfd85",
    journal = "Class. Quant. Grav.",
    volume = "38",
    number = "13",
    pages = "135014",
    year = "2021"
}

@article{Davis:2022ird,
    author = "Davis, D. and Littenberg, T. B. and Romero-Shaw, I. M. and Millhouse, M. and McIver, J. and Di Renzo, F. and Ashton, G.",
    title = "{Subtracting glitches from gravitational-wave detector data during the third LIGO-Virgo observing run}",
    eprint = "2207.03429",
    archivePrefix = "arXiv",
    primaryClass = "astro-ph.IM",
    reportNumber = "P2200192",
    doi = "10.1088/1361-6382/aca238",
    journal = "Class. Quant. Grav.",
    volume = "39",
    number = "24",
    pages = "245013",
    year = "2022"
}

@article{Davis:2022cmw,
    author = "Davis, Derek and Trevor, Max and Mozzon, Simone and Nuttall, Laura K.",
    title = "{Incorporating information from LIGO data quality streams into the PyCBC search for gravitational waves}",
    eprint = "2204.03091",
    archivePrefix = "arXiv",
    primaryClass = "gr-qc",
    reportNumber = "LIGO-P2200078",
    doi = "10.1103/PhysRevD.106.102006",
    journal = "Phys. Rev. D",
    volume = "106",
    number = "10",
    pages = "102006",
    year = "2022"
}

@article{Glanzer:2022avx,
    author = "Glanzer, J. and others",
    title = "{Data quality up to the third observing run of advanced LIGO: Gravity Spy glitch classifications}",
    eprint = "2208.12849",
    archivePrefix = "arXiv",
    primaryClass = "gr-qc",
    reportNumber = "LIGO-P2200238",
    doi = "10.1088/1361-6382/acb633",
    journal = "Class. Quant. Grav.",
    volume = "40",
    number = "6",
    pages = "065004",
    year = "2023"
}

@article{Glanzer:2023hzf,
    author = "Glanzer, Jane and Soni, Siddharth and Spoon, Jaidyn and Effler, Anamaria and Gonz{\'a}lez, Gabriela",
    title = "{Noise in the LIGO livingston gravitational wave observatory due to trains}",
    eprint = "2304.07477",
    archivePrefix = "arXiv",
    primaryClass = "astro-ph.IM",
    reportNumber = "LIGO-P2300032",
    doi = "10.1088/1361-6382/acf01f",
    journal = "Class. Quant. Grav.",
    volume = "40",
    number = "19",
    pages = "195015",
    year = "2023"
}

@article{Jackson:2020,
    author = "Jackson, C. and others",
    title = "{Teaching citizen scientists to categorize glitches using machine learning guided training}",
    doi = "10.1016/j.chb.2019.106198",
    journal = "Comput. Hum. Behav.",
    volume = "105",
    pages = "106198",
    year = "2020"
}

@article{Karki:2016pht,
    author = "Karki, S. and others",
    title = "{The Advanced LIGO Photon Calibrators}",
    eprint = "1608.05055",
    archivePrefix = "arXiv",
    primaryClass = "astro-ph.IM",
    doi = "10.1063/1.4967303",
    journal = "Rev. Sci. Instrum.",
    volume = "87",
    number = "11",
    pages = "114503",
    year = "2016"
}

@article{Kwok:2021zny,
    author = "Kwok, Jack Y. L. and Lo, Rico K. L. and Weinstein, Alan J. and Li, Tjonnie G. F.",
    title = "{Investigation of the effects of non-Gaussian noise transients and their mitigation in parameterized gravitational-wave tests of general relativity}",
    eprint = "2109.07642",
    archivePrefix = "arXiv",
    primaryClass = "gr-qc",
    doi = "10.1103/PhysRevD.105.024066",
    journal = "Phys. Rev. D",
    volume = "105",
    number = "2",
    pages = "024066",
    year = "2022"
}

@article{Macas:2022afm,
    author = "Macas, Ronaldas and others",
    title = "{Impact of noise transients on low latency gravitational-wave event localization}",
    eprint = "2202.00344",
    archivePrefix = "arXiv",
    primaryClass = "astro-ph.HE",
    doi = "10.1103/PhysRevD.105.103021",
    journal = "Phys. Rev. D",
    volume = "105",
    number = "10",
    pages = "103021",
    year = "2022"
}

@article{Malz:2025xdg,
    author = "Malz, Ann-Kristin and Veitch, John",
    title = "{Joint inference for gravitational wave signals and glitches using a data-informed glitch model}",
    eprint = "2505.00657",
    archivePrefix = "arXiv",
    primaryClass = "gr-qc",
    journal = "arXiv preprint",
    month = "5",
    year = "2025"
}

@article{Merritt:2021xwh,
    author = "Merritt, Jonathan and Farr, Ben and Hur, Rachel and Edelman, Bruce and Doctor, Zoheyr",
    title = "{Transient glitch mitigation in Advanced LIGO data}",
    eprint = "2108.12044",
    archivePrefix = "arXiv",
    primaryClass = "gr-qc",
    doi = "10.1103/PhysRevD.104.102004",
    journal = "Phys. Rev. D",
    volume = "104",
    number = "10",
    pages = "102004",
    year = "2021"
}

@article{Nguyen:2021ybi,
    author = "Nguyen, P. and others",
    title = "{Environmental Noise in Advanced LIGO Detectors}",
    collaboration = "LIGO Instrument Science Collaboration",
    eprint = "2101.09935",
    archivePrefix = "arXiv",
    primaryClass = "astro-ph.IM",
    doi = "10.1088/1361-6382/ac011a",
    journal = "Class. Quant. Grav.",
    volume = "38",
    number = "14",
    pages = "145001",
    year = "2021"
}

@article{Nuttall:2018xhi,
    author = "Nuttall, L. K.",
    title = "{Characterizing transient noise in the LIGO detectors}",
    eprint = "1804.07592",
    archivePrefix = "arXiv",
    primaryClass = "astro-ph.IM",
    doi = "10.1098/rsta.2017.0286",
    journal = "Phil. Trans. Roy. Soc. Lond. A",
    volume = "376",
    number = "2120",
    pages = "20170286",
    year = "2018"
}

@article{Ottaway:2012oce,
    author = "Ottaway, David J. and Fritschel, Peter and Waldman, Samuel J.",
    title = "{Impact of upconverted scattered light on advanced interferometric gravitational wave detectors}",
    doi = "10.1364/oe.20.008329",
    journal = "Opt. Express",
    volume = "20",
    number = "8",
    year = "2012"
}

@article{Pankow:2018qpo,
    author = "Pankow, Chris and others",
    title = "{Mitigation of the instrumental noise transient in gravitational-wave data surrounding GW170817}",
    eprint = "1808.03619",
    archivePrefix = "arXiv",
    primaryClass = "gr-qc",
    doi = "10.1103/PhysRevD.98.084016",
    journal = "Phys. Rev. D",
    volume = "98",
    number = "8",
    pages = "084016",
    year = "2018"
}

@article{Powell:2018csz,
    author = "Powell, Jade",
    title = "{Parameter Estimation and Model Selection of Gravitational Wave Signals Contaminated by Transient Detector Noise Glitches}",
    eprint = "1803.11346",
    archivePrefix = "arXiv",
    primaryClass = "astro-ph.IM",
    doi = "10.1088/1361-6382/aacf18",
    journal = "Class. Quant. Grav.",
    volume = "35",
    number = "15",
    pages = "155017",
    year = "2018"
}

@article{Robinet:2020lbf,
    author = "Robinet, Florent and Arnaud, Nicolas and Leroy, Nicolas and Lundgren, Andrew and Macleod, Duncan and McIver, Jessica",
    title = "{Omicron: a tool to characterize transient noise in gravitational-wave detectors}",
    eprint = "2007.11374",
    archivePrefix = "arXiv",
    primaryClass = "astro-ph.IM",
    doi = "10.1016/j.softx.2020.100620",
    journal = "SoftwareX",
    volume = "12",
    pages = "100620",    
    year = "2020"
}

@article{Smith:2011an,
    author = "Smith, Joshua R. and others",
    title = "{A Hierarchical method for vetoing noise transients in gravitational-wave detectors}",
    eprint = "1107.2948",
    archivePrefix = "arXiv",
    primaryClass = "gr-qc",
    reportNumber = "LIGO-DOCUMENT-P1100045",
    doi = "10.1088/0264-9381/28/23/235005",
    journal = "Class. Quant. Grav.",
    volume = "28",
    pages = "235005",
    year = "2011"
}

@article{Soni:2020rbu,
    author = "Soni, S. and others",
    collaboration = "LIGO Instrument Science Collaboration",
    title = "{Reducing scattered light in LIGO's third observing run}",
    eprint = "2007.14876",
    archivePrefix = "arXiv",
    primaryClass = "astro-ph.IM",
    reportNumber = "LIGO-P2000172",
    doi = "10.1088/1361-6382/abc906",
    journal = "Class. Quant. Grav.",
    volume = "38",
    number = "2",
    pages = "025016",
    year = "2020"
}

@article{Soni:2021cjy,
    author = "Soni, S. and others",
    title = "{Discovering features in gravitational-wave data through detector characterization, citizen science and machine learning}",
    eprint = "2103.12104",
    archivePrefix = "arXiv",
    primaryClass = "gr-qc",
    doi = "10.1088/1361-6382/ac1ccb",
    journal = "Class. Quant. Grav.",
    volume = "38",
    number = "16",
    pages = "195016",
    year = "2021"
}

@article{Soni:2023kqq,
    author = "Soni, Siddharth and Glanzer, Jane and Effler, Anamaria and Frolov, Valera and Gonz{\'a}lez, Gabriela and Pele, Arnaud and Schofield, Robert",
    title = "{Modeling and reduction of high frequency scatter noise at LIGO Livingston}",
    eprint = "2311.05730",
    archivePrefix = "arXiv",
    primaryClass = "astro-ph.IM",
    doi = "10.1088/1361-6382/ad494a",
    journal = "Class. Quant. Grav.",
    volume = "41",
    number = "13",
    pages = "135015",
    year = "2024"
}

@article{LIGO:2024kkz,
    author = "Soni, S. and others",
    collaboration = "LIGO Instrument Science Collaboration",
    title = "{LIGO Detector Characterization in the first half of the fourth Observing run}",
    eprint = "2409.02831",
    archivePrefix = "arXiv",
    primaryClass = "astro-ph.IM",
    doi = "10.1088/1361-6382/adc4b6",
    journal = "Class. Quant. Grav.",
    volume = "42",
    number = "8",
    pages = "085016",
    year = "2025"
}

\end{document}